\documentclass[11pt,twoside]{article}
\usepackage{asp2010}

\resetcounters

\begin{document}

\title{SONYC - Substellar Objects in Nearby Young Clusters}
\author{Koraljka~Mu\v{z}i\'c$^1$, Alexander~Scholz$^2$, Vincent~C.~Geers$^3$, Ray~Jayawardhana$^1$, and Motohide~Tamura$^4$
\affil{$^1$Department of Astronomy and Astrophysics, University of Toronto, 50 St. George St., Toronto ON M5S3H4, Canada}
\affil{$^2$Dublin Institute for Advanced Studies, 31 Fitzwilliam Place, Dublin 2, Ireland}
\affil{$^3$Institute f\"ur Astronomie, ETH, Wolfgang-Pauli-Strasse 27, 8093 Z\"urich, Switzerland}
\affil{$^4$National Astronomical Observatory of Japan, Osawa 2-21-2, Mitaka, Tokyo 181, Japan}}

\begin{abstract}

The origin of the lowest mass free-floating objects -- brown dwarfs 
and planetary-mass objects -- is one of the major unsolved problems in star  
formation. Establishing a census of young substellar objects is a  
fundamental prerequisite for distinguishing between competing  
theoretical scenarios. Such a census allows us to probe the initial  
mass function (IMF), binary statistics, and properties of accretion  
disks. Our SONYC (Substellar Objects in Nearby
Young Clusters) survey relies on extremely deep wide-field optical and  
near-infrared imaging, with follow-up spectroscopy, in combination  
with Spitzer photometry to probe the bottom end of the IMF to  
unprecedented levels. Here we present SONYC results for three  
different regions: NGC~1333, $\rho$ Ophiuchus 
and Chamaeleon-I. In NGC 1333, we  
find evidence for a possible cutoff in the mass function at 10-20  
Jupiter masses. In $\rho$ Oph we report a new brown
dwarf with a mass close to the deuterium-burning limit.
\end{abstract}

\section{Introduction}
\label{intro}
The origin of the stellar initial mass function (IMF) is one of the major
issues in astrophysics. The low-mass end of the IMF, in particular, has been subject of
numerous observational and theoretical studies over the past decade (see \citealt{bonnell07}).
Star forming regions and young clusters harbor a large population of free-floating objects below
the stellar-mass boundary, found to be at $\sim$0.075$\,M_{\odot}$. These objects include 
brown dwarfs (BDs), but also a population of objects with masses comparable to those of massive planets
(we refer to these objects as ``planetary-mass objects'' or PMOs). 
Despite the large advances in our understanding of substellar objects over the past decade, some of the
most important questions still remain unanswered. The origin of BDs and PMOs is not clear \citep{whitworth05};
competing scenarios include turbulent fragmentation \citep{padoan&norlund04}, ejection from
multiple systems \citep{bate09}, and ejection from fragmenting protoplanetary disks \citep{stamatellos08}.
The shape of the IMF at very low
masses is the subject of an ongoing debate in the literature \citep{bonnell07,chabrier03}.
In nearby star forming regions the total number BDs relative to the number of low-mass stars varies
between 3 and 8 with large uncertainties \citep{andersen08}. The number
of PMOs and its dependence on environment is even more uncertain. The IMF could be still rising
below 0.015$M_{\odot}$ \citep{caballero07}, or declining in this regime \citep{lucas05}. 
A cutoff in the mass function has not been observed yet. 

SONYC -- Substellar Objects in Nearby Young Clusters -- is an ongoing project to
provide a complete census of the brown dwarf and planetary mass object population
in nearby young clusters, and to establish the frequency of substellar mass objects as a
function of cluster environment. The resulting catalog of substellar mass candidates will
provide the basis for detailed characterization of their physical properties (disks, binarity,
atmospheres, accretion, activity). 
The primary means of identifying candidates is broad-band imaging in the optical and the infrared, 
thus aiming to detect the photosphere. 
The survey is also combined with the 2MASS and Spitzer photometry catalogs. Photometric selection
results in large samples of candidates and requires extensive spectroscopic follow-up to asses the real
nature of the objects. 
Our observations are designed to
reach limiting masses of $\sim\,$0.005$\,M_{\odot}$, 
well below the deuterium-burning limit at 0.015$\,M_{\odot}$, and thus require us of 4- to 8-m-class telescopes.
By probing several star forming regions we want to probe for environmental differences in the frequency and properties
of substellar objects.

In this contribution, we summarize the results delivered in the framework of SONYC over the past two years.
We have surveyed three star forming regions: NGC~1333 \citep{scholz09}, $\rho$~Ophiuchus \citep{geers10}, and
Chamaeleon-I (Mu\v{z}i\'c et al., submitted to ApJ).

\section{NGC~1333}
\label{N1333}
\begin{figure}[!ht]
\plotone{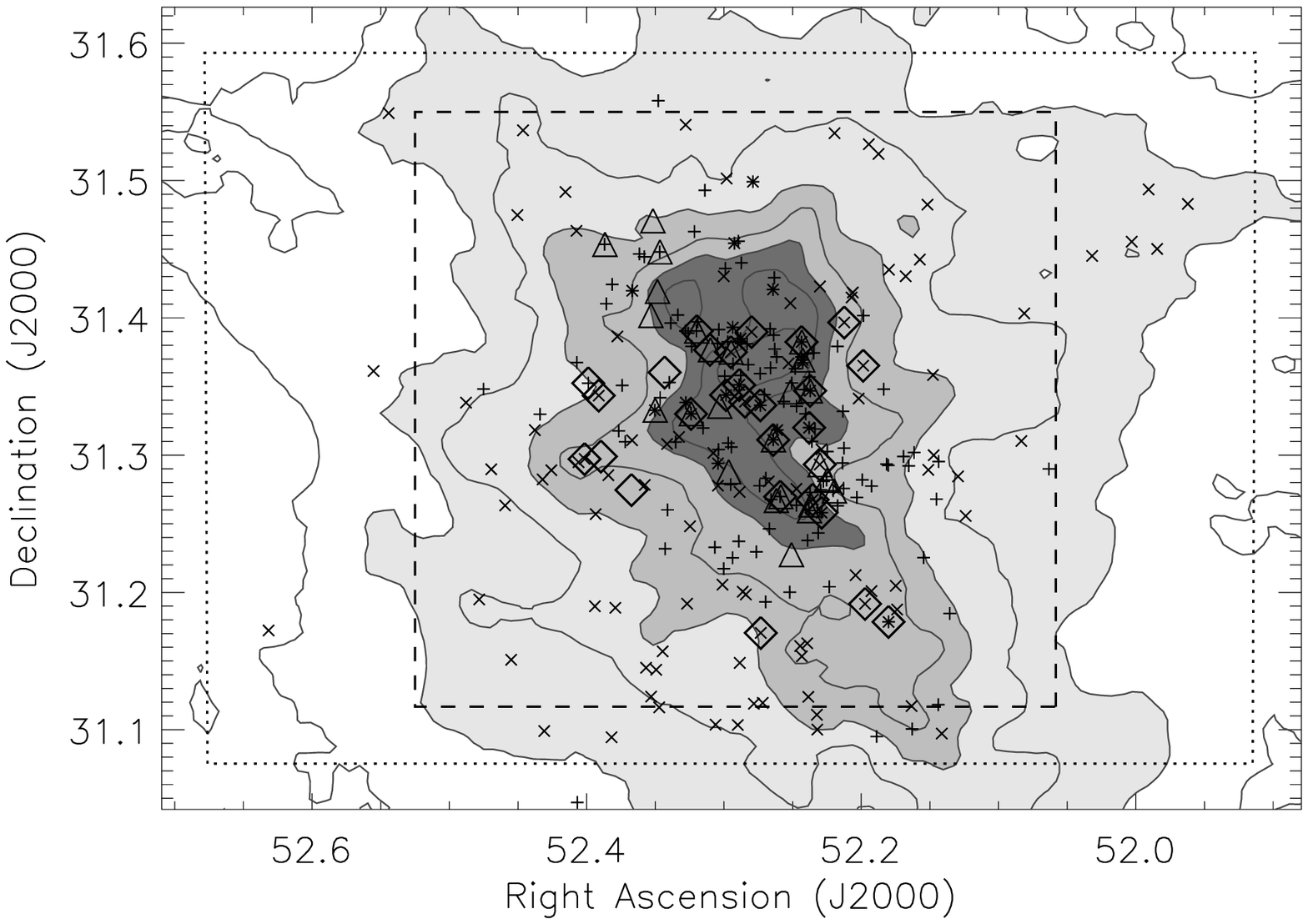}
\caption{Spatial coverage of our survey in NGC~1333. Crosses are all candidates selected from optical photometry 
with $i'>$18.5, diamonds the spectroscopically confirmed substellar members, triangles the confirmed BDs from 
\citet{wilking04}. The plusses are YSO candidates from Spitzer data without spectroscopic confirmation from 
\citet{gutermuth08}. The underlying contours show the distribution of $^{13}$CO from the $J$=1 -- 0 map by 
\citet{ridge06}. The dotted and dashed lines show the area covered in our optical and NIR survey, respectively.
Image from \citet{scholz09}; reproduced by permission of the AAS.}
\label{N1333sd}
\end{figure}

NGC~1333 is a cluster in the Perseus star forming complex, with an age of $\sim$1$\,$Myr, a distance of 
$\sim$300$\,$pc \citep{dezeeuw99,belikov02}, and moderate extinction. The spatial coverage of our survey in NGC~1333 is 
0.25$\,$deg$^2$ (see Fig~\ref{N1333sd}).

\subsection{Photometry and Spectroscopy}
\label{N1333_ps}
Imaging of NGC~1333 was performed at the Subaru telescope. We used the Suprime-Cam wide-field
optical imager \citep{miyazaki02} to obtain images in SDSS filters $i'$ and $z'$, and MOIRCS \citep{suzuki08,ichikawa06}
for near-infrared $J$ and $K_S$ observations. Our data are complete down to $i'$=24.7, 
$z'$=23.8, $J$=20.8, and $K_S$=18.0 mag. In terms of object masses
for members of NGC~1333, this roughly corresponds to mass limits of 0.008$M_{\odot}$ for $A_V\lesssim$10 
and 0.004$M_{\odot}$ for $A_V\lesssim$5, based on the COND03 \citep{baraffe03} and DUSTY00 \citep{chabrier00} 
evolutionary tracks.
We used MOIRCS again to carry out multi-object spectroscopy for 53 sources in NGC~1333, selected on the basis
of their broad-band colors that are in the range as expected for substellar cluster members. This sample
does not show any bias in spatial coverage or optical/NIR colors with respect to the full photometric
candidate sample. The wavelength coverage for MOIRCS low-resolution spectroscopy includes $H$- and $K$-bands. 

\subsection{Results}
\label{N1333_res}

\begin{figure}[!ht]
\plotone{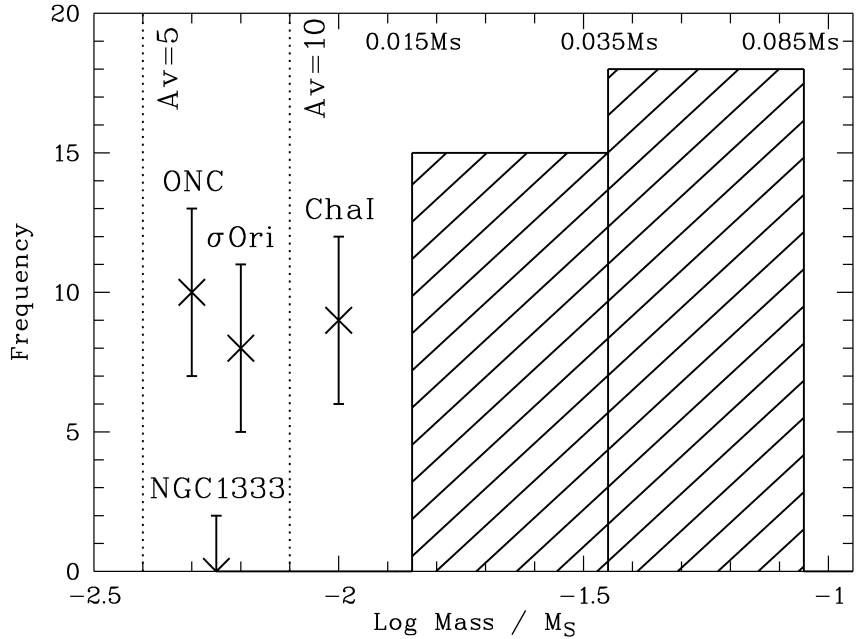}
\caption{Mass distribution of BDs in NGC~1333 (hatched histogram) and the deficit of planetary-mass objects. The three
labeled data points show the predicted number of PMOs in NGC~1333 based on the surveys in $\sigma$~Ori, 
the ONC, and in Cha-I. The error bars correspond to 1$\sigma$.
Image from \citet{scholz09}; reproduced by permission of the AAS.}
\label{N1333massd}
\end{figure}

Our goal is to identify young sources with effective temperatures at or below 3000$\,$K. As outlined
in detail in \citet{scholz09}, these objects show a characteristic spectral shape in the
near-infrared. In particular, their spectra have a clear peak in the H-band \citep{cushing05}. 
This feature is caused by water absorption on both sides of the H-band. The depth
of this feature depends strongly on effective temperature. While the H-band peak appears
round in old field dwarfs, it is triangular in young, low-gravity sources. 
In addition, young brown dwarfs are expected to have flat or increasing $K_S$-band
spectra with CO absorption bands at $\lambda>\,$2.3$\,\micron$.
To estimate the effective temperatures of the BD candidates we perform fitting of
model spectra from the DUSTY series \citep{allard01} to our observed spectra. We confirm 19
sources with effective temperatures of 3000~K or lower. Combined with the clear indications
for youth, we classify them as probable substellar members of NGC~1333. 

Combining our survey results with previous studies, the
current census of spectroscopically confirmed BDs in
NGC~1333 is 33. The ratio of substellar to stellar members with
masses below 1$M_{\odot}$ is 1.5$\,\pm\,$0.3, lower by a factor of 2 -- 5
than in all other previously surveyed regions. Thus, NGC~1333 clearly
shows an overabundance of BDs. On the other hand, the cluster shows lack
of PMOs. The low-mass limit of the confirmed BDs
is 0.012 -- 0.02$M_{\odot}$, but the completeness limits are at significantly
lower masses. Scaling from literature results in other regions
($\sigma$ Ori, ONC, Cha-I), we would expect to find 8 -- 10
PMOs, but we find none (see Fig~\ref{N1333massd}). This indicates a cutoff in
the mass spectrum around the deuterium-burning limit in
NGC~1333.  

\section{$\rho$ Ophiuchus}
$\rho$~Oph is one of the closest (d = 125$\,\pm\,$25 pc, \citealt{degeus89}) regions of active star
formation. $\rho$~Oph cluster is not
as compact as the first SONYC target NGC~1333, 
and exhibits extremely high and variable levels of extinction. The main cloud,
L1688, is a dense molecular core, with visual extinction up to 50 -- 100 mag \citep{wilking&lada83}, 
hosting an embedded infrared cluster of around 200 stars, inferred to have a median
age of 0.3 Myr, and surrounded by multiple clusters of young stars with a median age of 2.1
Myr (\citealt{wilking05} and references therein).

\subsection{Photometry and spectroscopy}
\begin{figure}[!ht]
\plotone{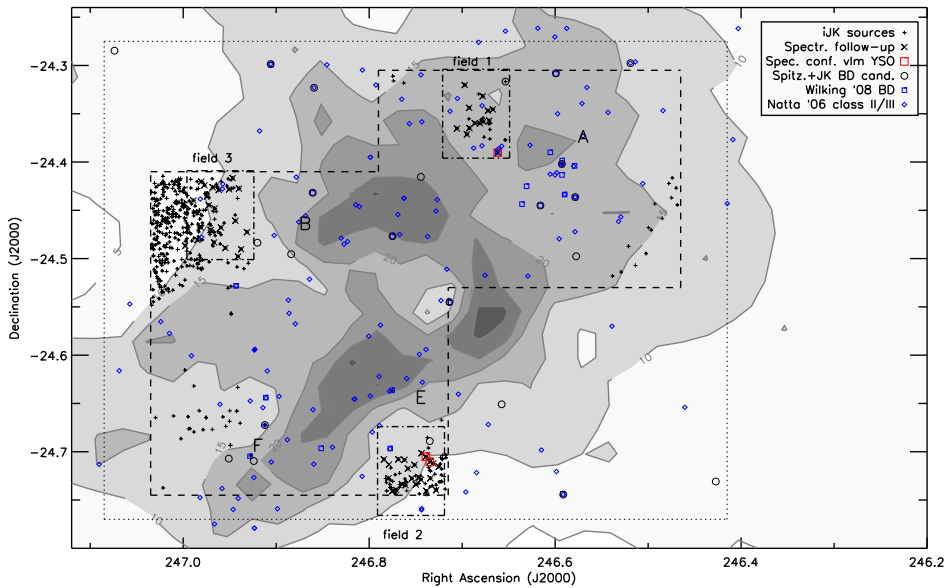}
\caption{
Spatial distribution of sources in $\rho$~Oph. Contours are $A_V$ = 5, 10, 15, 20, 25, 30,
as derived from 2MASS by the COMPLETE project. Dotted line: $i$-band imaging coverage;
dashed line: $J$ and $K_S$-band imaging coverage; dash-dot line: MOIRCS spectroscopy fields.
$iJK_S$ catalog sources indicated with +; $iJK_S$ catalog selected BD candidates with follow-up
MOIRCS spectroscopy indicated with x; spectroscopically confirmed low mass YSOs 
are indicated with red large squares. Candidate BDs selected from $JK_S$
+ Spitzer photometry indicated with circles. Previously known BDs from \citet{wilking08} 
and class II / III sources from \citet{natta06} are indicated with small blue
squares and diamonds respectively.
Figure from \citet{geers10}; reproduced by permission of the AAS.}
\label{rhoophsp}
\end{figure}

For $\rho$~Oph we used the same instrumental setup as for NGC~1333 (Section~\ref{N1333_ps}), to
obtain broad-band images in $i'$, $J$ and $K_S$, and $HK$ spectroscopy.
Completeness limits are placed at $i'$=24.2, $J$=20.6, and $K_S$=17.8. This corresponds to
mass limits of 0.004 -- 0.1$M_{\odot}$ in the $i'$-band, and 0.001 -- 0.007$M_{\odot}$ in the $J$-band, 
for extinction of $A_V$ = 5 -- 15 (based on the COND03
and DUSTY00 evolutionary tracks).
Photometric catalogs were cross-correlated with the existing Spitzer data.
The spatial coverage of our $i'$+$JK_S$ survey
is 0.171 deg$^2$ (see Fig~\ref{rhoophsp}).

\begin{figure}[!ht]
\plotone{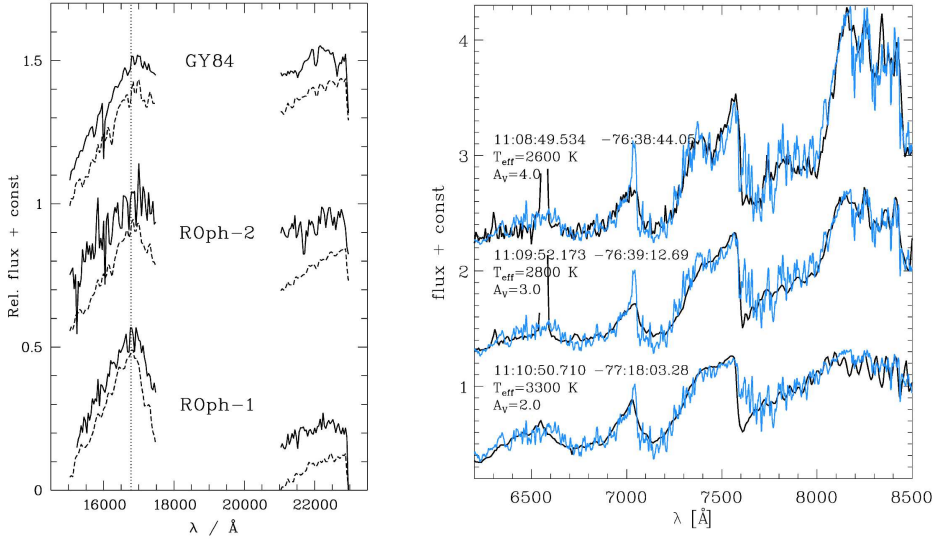}
\caption{Left panel: MOIRCS spectra (solid) of candidate substellar objects in $\rho$~Oph, with best fit
reddened model (dashed) with an $T_{\mathrm{eff}}$ of 2500 K, 3100 K, and 3400 K; offsets were applied
for clarity. 
Figure from \citet{geers10}; reproduced by permission of the AAS.
Right panel: Examples of the model fitting for some of the previously known
Cha-I candidates. In black are shown the spectra obtained by VIMOS/VLT, with the best-fit models in blue (AMES-Dusty; \citealt{allard01}).  
}
\label{spectra}
\end{figure}

From the optical and near-infrared photometry, 309 objects were selected as candidate
substellar cluster members. 58 of these objects, and 1 additional
previously known BD candidate, were targeted for follow-up spectroscopy. Based on
multi-object spectroscopy, using the water absorption features
in the H-band, 1 of the 58 new candidates was confirmed as a substellar mass object
with $T_{\mathrm{eff}}$ = 2500 K (see left panel in Fig~\ref{spectra}). 
From MOIRCS, 2MASS, and Spitzer photometry a sample of 27 sources with mid-infrared
color excess and near-infrared colors indicative for substellar mass sources
with disks are identified. Of these, 11 are previously spectroscopically confirmed brown
dwarfs, while 16 are newly identified candidates.
Based on present day surveys of the stellar and brown dwarf populations, the ratio of
substellar to stellar sources in $\rho$~Oph is derived to have an upper limit of 5 -- 7, 
in line with other nearby young star forming regions. 
Census of substellar objects in $\rho$~Oph, based on current existing surveys, 
is likely highly incomplete, due to the variable
and high extinction.

\section{Chamaeleon I}
At a distance of $\sim\,$160$\,$pc \citep{whittet97}, Chamaeleon I (hereafter Cha-I) harbors a rich population of known YSOs \citep{luhman07}. Compared to the other two SONYC targets, 
Cha-I is less compact than NGC1333 and has less extinction than $\rho$~Oph. The spatial coverage of our survey 
in Cha-I is 0.25$\,$deg$^2$.

\subsection{Photometry and spectroscopy}
This study is primarily based on data from the European Southern Observatory, obtained in runs 078.C-0049 and 382.C-0174. The first campaign provided deep optical $iz$ imaging (VIMOS/VLT; \citealt{lefavre03}), and near-infrared $JK_S$ imaging 
(SOFI/NTT; \citealt{moorwood98}). In our imaging survey, we reach completeness limits of 23.0 in $I$, 18.3 in $J$, and 16.7 in $K_S$. This corresponds 
to mass limits of $(0.002 - 0.01) M_{\odot}$, for $A_V=0-10$, based on 
DUSTY00 and COND03 models. From the optical dataset we selected a list of candidate members which were observed using multi-object spectrograph VIMOS/VLT, using the low resolution red grism (5500 -- 9000 \AA). Additionally, we made use of the archival Spitzer images for our target regions. 

\subsection{Results}

From the optical photometry, 142 objects were selected as candidate
substellar cluster members. 60 of these objects have been observed in the spectroscopy campaign, however, due to the 
observed limit for spectroscopy at I$\approx$21, only 18 of these candidates have spectra that can be used for
classification. In addition, we obtained spectra for more than 200 randomly selected objects in the VIMOS field-of-view.
Red portion of M-dwarf spectra is dominated by molecular features \citep{kirkpatrick91,kirkpatrick95}, which
allows relatively simple preliminary selection of candidates based on visual inspection. 
To determine the effective temperature of the remaining candidate objects we performed spectral fitting using the AMES-dusty models \citep{allard01}. We identify 13 objects consistent with the spectral type M, of which 11 are previously known M-dwarfs with confirmed membership in the cluster (see right panel in Fig~\ref{spectra}). The two newly reported objects have effective temperatures consistent with masses above the substellar limit. 

Based on the results of our survey and combined with the numbers of substellar objects from the literature, we estimate that the number of the missing low-mass members down to $\sim 0.008\,M_{\odot}$ for $A_V\leq5$ in Cha-I is $\leq\,$7, i.e. $\leq3\%$ of the total number of members according to the current census. We might, however, still miss objects with lower masses, and objects at higher extinctions.

\section{Conclusions}
It is clear that the census of BDs and PMOs in most star forming regions is still incomplete. 
Based on the existing data we can conclude that: (a) there are hints of regional differences in the mass function
at the very-low-mass end, and (b) only a combination of different search techniques can provide a robust picture  
of the substellar population. 

\acknowledgements 
The research was supported in part by grants from the Natural
Sciences and Engineering Research Council (NSERC) of Canada to RJ.
This work was supported in parts by the Science Foundation Ireland  
within the Research Frontiers Programme under grant no. 10/RFP/AST2780.
MT is supported by a Grant-in-Aid for Specially Promoted Research and by the Mitsubishi Foundation.


\end{document}